\documentclass[prl,twocolumn,showpacs]{revtex4}
\usepackage{graphicx,amsmath}
\usepackage{txfonts}
\begin{document}
\title{
On a relationship between the collective migration of surface 
atoms in microclusters and the saddle points on the potential energy surface
}
\author{Yasushi Shimizu\thanks{e--mail:
shimizu@ifc.or.jp}}
\address{
Fukui Institute for Fundamental Chemistry, Kyoto University, 
34-4, Takano Nishi-hiraki-cho, Kyoto 606-8803, Japan
}
\author{Shin-ichi Sawada}
\address{
School of Science and Technology, Kwansei Gakuin University,
2-1, Gakuen, Sanda 669-1337, Japan
}
\author{Kensuke S.Ikeda }
\address{
Department of Physics, 
Ritsumeikan University, Noji-higashi 1-1-1, Kusatsu 525-8577, Japan
}
\date{Nov 24, 2002}
%-------------------------------------------------------------------------
\begin{abstract}
%%%%%%%%%%%%%%%%%%%%%%%%%%%%%%%%%%%%%%%%%%%%%%%%%%%%%%%%%%%%%%
Plenty of saddles on a multidimensional potential energy surface(PES) of 
two-dimensional microclusters, where atoms are interacting via 
Morse potential, are numerically located. 
The reaction paths emanating from the two types of the local minima, which represent the 
compact and the non-compact shape of Morse clusters, to their neighboring 
saddles on PES are elucidated.
By associating the reaction path crossing these saddles with the atomic 
rearrangements,
we evaluate the barrier height corresponding to various characteristic 
atomic motion accompanied by the {\it floaters} 
(i.e. surface atoms popped out of the cluster surface).
Our findings are summarized as: 
(i)The saddle points implying the {\it gliding motion} 
of a single {\it floater} over the cluster surface yields 
extremely small values of the  energy barriers 
regardless of the shapes of clusters. 
In particular, the {\it gliding motion} of a train composed of a 
few surface atoms also appears as the low-lying saddles. 
As a result, the barrier height corresponding to the  
{\it simultaneous gliding motion}, which is a manifestation of the 
reaction path crossing the higher-index saddles on PES, is significantly low.
(ii)A surface rearrangement, where {\it floaters} are 
created or annihilated,   
%accompanied with not only {\it floaters} 
%but also atoms close to the surface, 
implies relatively high barrier energy which is still accessible 
below melting point. 
(iii)On the other hand, the atomic motion, where atoms located deep inside of 
the clusters are rearranged as well as surface atoms, yields 
extremely high barrier energies. 
Some relations between these results and the recent experimental study 
of the surface cluster diffusion are also pointed out.
\end{abstract}
%\pacs{05.60.-k, 36.40.Sx, 61.46.+w, 66.30.Jt, 67.80.Mg}
%%%%%%%%%%%%%%%%%%%%%%%%%%%%%%%%%%%%%%%%%%%%%%%%%%%%%%%%%%%%%%%%%%%%%%%
\maketitle
%{\it Introduction}-
One of the most remarkable dynamical features of microclusters which are
experimentally and theoretically observed is that their 
motion is dominated by large fluctuation\cite{sugano,iijima}.
The floppy motion of microclusters can be attributed to a wandering motion 
among many local minima which are partitioned by substantially low
saddles on the multidimensional potential energy surface(PES)
\cite{ajayan,sawada}.
In fact a PES of atomic clusters is populated by many saddles and
minima even if a cluster contains at most order of $10^1$ atoms.
A number of minima and saddles of microclusters are enumerated by brute
force of the current computational power, while a variety of schemes to 
search for the saddle point is proposed. 
Provided that the size of cluster is enough small, it is possible to
obtain nearly complete distributions of minima and saddles.  
An accumulation of such numerical research has been shed 
light into the global feature of potential landscape 
whose topography is considered to be a origin of 
the complicated and exotic nature of the glass forming material, 
microclusters, peptides and proteins\cite{wales}. 
If we restrict ourselves to microclusters, there exists a wide variety of 
characteristic atomic motion in it as well as that on a bulk surface. 
A representative example is motion of a {\it floater} which continues 
to wander among stable sites on the cluster surface. 
According to Cheng and Berry, the {\it floaters} are surface atoms which 
are popped out of the surface and keep migrating above or outside the 
outer layer of the cluster as shown in Fig.1\cite{berry_floater}.
%%%%%%%%%%%%%%%%%%%%%%%%%%%%%%%%%%%%%%%%%%%%%%%%%%%%
%  Fig.1
%%%%%%%%%%%%%%%%%%%%%%%%%%%%%%%%%%%%%%%%%%%%%%%%%%%%
\begin{figure}[htbp]
\begin{center}
\resizebox{5cm}{!}{\includegraphics{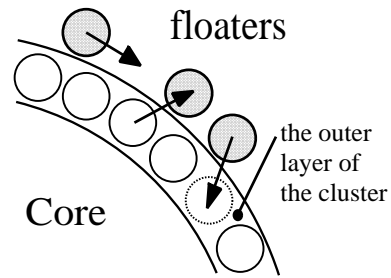}}
\caption{\label{fig:fig1} \footnotesize
A schematic picture of creation and annihilation of the floaters in the 
vicinity of the cluster surface. Surface atoms are denoted by white circles, while 
floaters are indicated by shaded circles. A vacancy is indicated by a 
dotted circle.}
\end{center}
\end{figure}
%%%%%%%%%%%%%%%%%%%%%%%%%%%%%%%%%%%%%%%%%%%%%%%%%%%%
Since the majority of constituent atoms of a cluster are located
on surface, many {\it floaters} are expected to be created and 
annihilated throughout dynamics. 
Most of the previous works related to the atomic diffusion 
on a surface put their focus on the motion of a single surface atom.
However, these {\it floaters} may interact with each other and move 
collectively. 
A collective migrating motion of floaters are expected to 
play an important role in the mass-transport 
such like an alloying behavior of the nano-sized metal clusters 
and an island migration on a bulk surface\cite{yasuda,fink,voter}. 
Indeed, the migration of atoms along the edge of a surface
Ar and Ir cluster was experimentaly observed\cite{wen,wang}. 
Moreover, it was numerically verified that the active surface 
atoms move freely and cooperatively along the cluster surface and
diffuse into the inside of a cluster due to an accumulation of 
rearrangements of surface atoms well below the 
melting point\cite{sis}.
On the other hand, a detail analysis of the collective motion of surface
atoms are still left untouched.  
The most straightforward way probing a complicated motion of surface
atoms is to clarify the relationship between 
the characteristic motion of surface atoms in a cluster 
and the corresponding reaction path on a multidimensional PES\cite{voter}. 
%%%%%%%%%%%%%%%%%%%%%%%%%%%%%%%
%purpose
%%%%%%%%%%%%%%%%%%%%%%%%%%%%%%%
The purpose of this Letter is to give a brief sketch of a
relationship between characteristic motion of surface atoms in a 
configurational space and that on 
a multidimensional PES by putting a special emphasis on the
reaction paths connecting a local minimum and its neighboring saddles 
which are climbed over during an isomerization process. 
%Information about the saddle points is practically of importance, 
%because the barrier height can be directly 
%associated with quantities such as activation energy of the 
%diffusion coefficient by assuming the transition state theory\cite{vineyard}. 
\par
%%%%%%%%%%%%%%%%%%%%%%%%%%%%%%%%%%%%%%%
We employ two-dimensional(2D) Morse model given by $U=\sum_{i<j} V(r_{ij})$, where 
the pair potential function $V$ is given by
$V(r)=\epsilon \{ e^{-2\beta (r-r^c)}- 2 e^{-\beta(r-r^c)} \}.$
Note that $r$ is the separation between two atoms. 
We choose $\beta=1.3588[A^{-1}]$, $\epsilon=0.3429[eV]$ and
$r^c=2.866[A]$, those  are suitable values for copper\cite{Giri}.
The reasons why we select a 2D system are two folds:
One is that the present model is a concise model of the atomic
rearrangements in an island on bulk surface, 
if one neglects the interaction between a cluster and substrate. 
As the temporal motion of atoms in 2D cluster was recently observed by the field 
ion microscope(FIM)\cite{wen,wang},  
a close analysis of the present 2D model is expected to 
bring some insights into these experimental results as mentioned below. 
%The relation to these experiments are mentioned below. 
Another reason is that it is an illustrative and  transparent example
which demonstrates how cluster atoms move individually and collectively 
in a microcluster at a glance. 
In addition, most of the results obtained in the present 2D model 
are expected to remain unaltered in the 
realistic 3D model as far as the following topics are concerned. 
A numerical search for the saddle points which are reachable from the given 
local minimum is carried out in terms of the so-called eigenvector-following method
\cite{cerjan}. 
The saddle points are explored by starting the searching procedure 
from a given local minimum on PES.
More precisely, a random fluctuation is added to the atomic
configurations at a local minima in order to reach every saddle point  
which may be randomly distributed about the given local minima. 
The reachable saddle points are enumerated for 10000 initial points. 
The intensity of random fluctuation is about $7\%$ of the mean
separation of the neighboring atoms. 
Although no one guarantees that all of the saddle points connected to 
a given local minimum are completely detected by the present method, the authors
confirmed that almost all trivial saddles which are noticeable by 
inspection are successfully found in various size and shape of Morse
clusters. The present method is applied to pick up the distinct saddle points of
the Morse clusters $M_{67}$. 
The size of the cluster is chosen to give enough number of the saddles points 
to extract a significant statistical trend from data.  
In particular, two kinds of the clusters different in shape are examined. 
One is a compact cluster i.e., a geometrically packed structure, 
which is located at a deep minimum on PES. 
Another is a non-compact cluster which is possessed by 4 {\it floaters} 
capable to move almost freely, as shown in Fig.3 and 4\cite{note1}.
The presence of many floppy {\it floaters} is an indication of the fact
that unpacked configurations of a cluster is located at a shallow 
minimum on PES. 
The resulting number of the numerically found 1st and 2nd order saddles 
of the non-compact $M_{67}$ are 89  and  656, respectively. 
Those for the compact $M_{67}$ are 185  and 1020, respectively. \par
%%%%%%%%%%%%%%%%%%%%%%%%%%%%%%%%%%%%%%%%%%%%%%%%%%
%  Fig.2
%%%%%%%%%%%%%%%%%%%%%%%%%%%%%%%%%%%%%%%%%%%%%%%%%%
%%%%%%%%%%%%%%%%%%%%%%%%%%%%%%
%
%%%%%%%%%%%%%%%%%%%%%%%%%%%%%%
%%%%%%%%%%%%%%%%%%%%%%%%%%%%%%%%%%%%%%%%%%%%%%%%%%%%%%%%%%
In Fig.2 the number of the 1st and 2nd order saddles with respect to the value 
of the barrier height is displayed for the compact and the non-compact $M_{67}$. 
%%%%%%%%%%%%%%%%%%%%%%%%%%%%%%%%%%%%%%%%%%%
%  Fig.2
%%%%%%%%%%%%%%%%%%%%%%%%%%%%%%%%%%%%%%%%%%%
\begin{figure}[htbp]
\begin{center}
\resizebox{8.5cm}{!}{\includegraphics{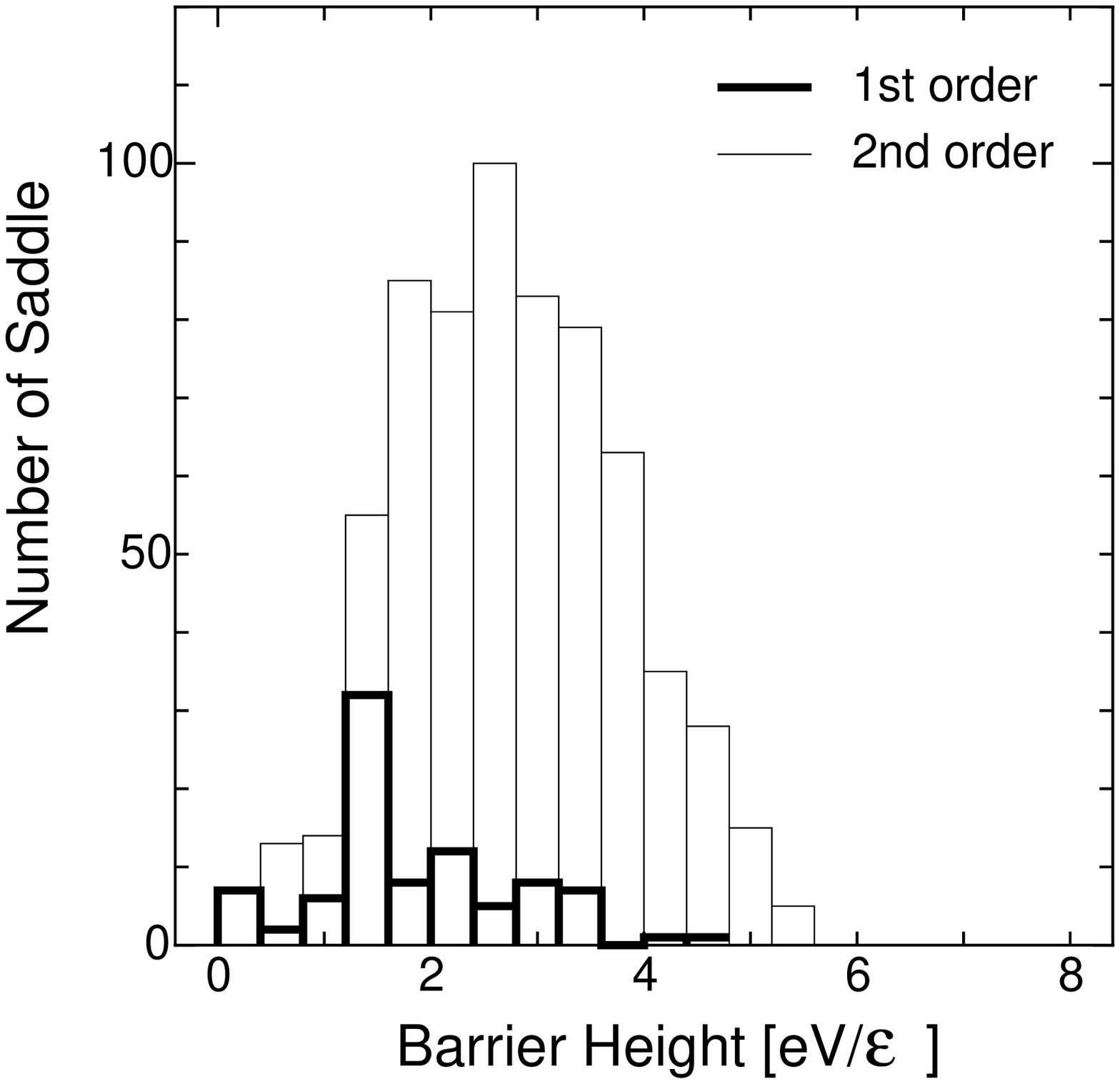}
\includegraphics{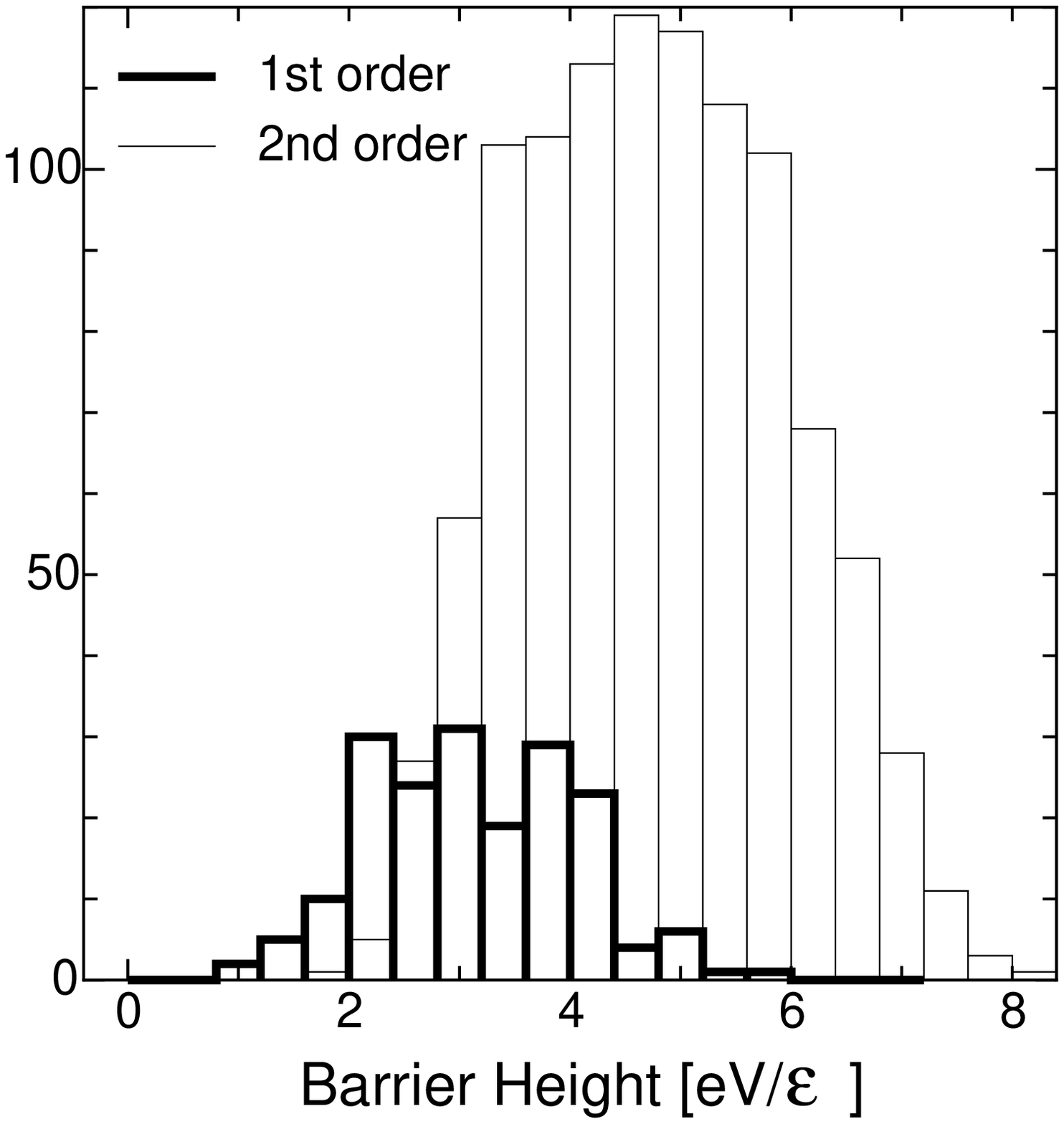}}
\caption{
Histograms for the number of the saddle points with respect to 
the value of the barrier height. The barrier energy is normalized by 
the depth of the pair potential $\epsilon$.
(a)The data of the 1st and 2nd order index saddles for the non-compact $M_{67}$ are shown
 by thick and thin line, respectively.
(b)The data of the 1st and 2nd order index saddles for the compact 
$M_{67}$ are also shown by thick and thin line.
\label{fig:fig2}}
\end{center}
\end{figure}
%%%%%%%%%%%%%%%%%%%%%%%%%%%%%%%%%%%%%%%%%%%
One can immediately notice that the non-compact cluster 
has more saddles 
lying in the low energy region both for the 1st and 2nd order index saddles.
If one takes into account the presence of many floppy {\it floaters},  
it is not a surprising result to appear more low-lying barriers 
near the local minimum representing the non-compact $M_{67}$. 
Indeed, as exhibited in Fig.3(a), the reaction paths crossing the extremely
low-lying barrier, whose height is about $0.3\epsilon$, are attributed to 
the hopping motion of a single {\it floater}. 
Similarly, the bounded train of a few surface atoms is easy 
to glide over the cluster surface as shown in 
Fig.3(b) and Fig4(g), and the barrier heights for
such motion are considerably low (about $0.6\epsilon$ and $0.9\epsilon$, respectively).
Since a hopping of a floater costs about $0.3\epsilon$, 
the barrier height for the gliding motion shown in Fig.3(b) and Fig.4(g)
are roughly estimated as $0.6\epsilon(0.3\epsilon\times 2)$ and 
$0.9\epsilon(0.3\epsilon\times 3)$, respectively. 
The migrating motion of a bounded train composed of the surface 
atoms running along the edge is experimentally observed in the 2D cluster of 
$Ir_{18}$ and $Ir_{36}$, and is a possible elementary process of the 
periphery diffusion, which is one of the driving factors for the motion
of a 2D $Ir$ cluster on $Ir(111)$\cite{wang,metiu}.
%\par
%%%%%%%%%%%%%%%%%%%%%%%%%%%%%%%%%%%%%%%%%%%
%  Fig.3
%%%%%%%%%%%%%%%%%%%%%%%%%%%%%%%%%%%%%%%%%%%
\begin{figure}[htbp]
\begin{center}
\resizebox{9.6cm}{!}{\includegraphics{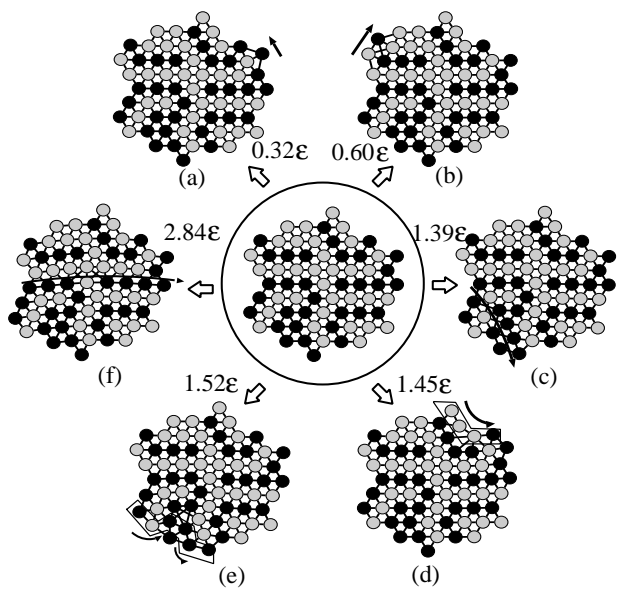}}
\caption{
Representative configurations of the cluster atoms corresponding to the saddle points for 
the non-compact $M_{67}$ are exhibited by balls and sticks. 
Sticks are inserted between pair atoms whose separation is shorter than $1.5r^c$.
In the center circle, the configuration of the 
non-compact $M_{67}$ at a local minima is displayed.
Typical atomic configurations of the 1st order saddle points are shown 
in (a)-(f). 
The direction of the atomic displacement from the initial local minimum to 
another beyond the saddle point is denoted by arrows.
The barrier height is also inserted in the figure.
Each atom is colored by black or gray randomly just to identify its 
location before and after the displacement.
\label{fig:fig3}}
\end{center}
\end{figure}
%%%%%%%%%%%%%%%%%%%%%%%%%%%%%%%%%%%%%%%%%%%
On the other hand, it should be noted that there appear significantly many
low-lying 2nd order saddles on the PES for the non-compact $M_{67}$.  
In Fig.5 the atomic configurations of the typical low-lying higher-index saddle points
are depicted. Those low-lying saddles reveal the {\it simultaneous gliding motion} 
occurring at spatially distant places on a cluster surface, whilst the
generic motions yielding higher-order index saddles are not 
necessarily decomposed into such a {\it almost independent} gliding motion.
%For the present case the higher-index saddles appear in a considerably low energy region. 
In contrast, in a usual molecular reaction, the barrier of the higher-index saddles 
are so high that the reaction paths crossing them are neglected.
As is evident from the barrier energy indicated in Fig.5, these
higher-index saddles are easily accessible even if the temperature of the system is 
well below the melting point\cite{note2}.
The presense of many low-lying higher-index saddles is a remarkable
ingredient of a medium-sized cluster. 
If the given cluster shape located at a local minimum on PES 
is connected with $k$ saddles of the low-lying 1st order index,   
the number of the low-lying $l$-th order index saddle which is 
decomposable into the 1st order saddle is about $_kC_l$, the binomial coefficient for $l$
items selected from $k$. 
Although microclusters tend to have such low-lying higher-index saddles 
irrespective of the cluster shapes as shown in Fig.5, the discrepancy 
in Fig.2 implies that the presence of many floaters significantly reduces 
the barrier height implying these higher-index saddles.
%\par
%%%%%%%%%%%%%%%%%%%%%%%%%%%%%%%%%%%%%%%%%%%%%%%%%%
Moreover, it is important to note that 
the surface atoms other than {\it floaters} also play an 
crucial role in a surface rearrangement, where a {\it floater} is newly 
created or annihilated as demonstrated in 
Fig.3(c)-(e) and Fig.4(h)-(k). 
The barrier energies for these motions are 
slightly larger than those for a simple {gliding} motion of {\it
floaters} or surface atoms forming a train. 
In the crudest approximation, the resultant barrier height is determined 
by the number of atoms involved by the displacement. 
The cost for a displacement which undergoes a single
bond breaking is about $1.0\epsilon$, while a hopping of a floater costs
about $0.3\epsilon$. 
Thus, we can estimate that the displacements  
in Fig.3(c), (d), and (e) cost about $1.2\epsilon(0.3\epsilon\times 4)$, 
$1.2\epsilon(0.3\epsilon\times 4)$ and 
$1.8\epsilon(0.3\epsilon\times 6)$, respectively.
Similarly, the displacements given in Fig.4(h), (i),(j), and (k) demand  
about $2.0\epsilon (1.0\epsilon \times 2)$, $2.0\epsilon (1.0\epsilon \times
2)$,  $2.8\epsilon (0.3\epsilon \times 6+1.0\epsilon)$, and
$2.2\epsilon (0.3\epsilon \times 4+1.0\epsilon )$, respectively. 
%\par
As demonstrated in these examples, there exits a wide variety of
collective motion accompanied with surface atoms and {\it floaters}.
Such collective behaviors are experimentally of interest. 
Considering the present Morse cluster as a model of an island on a bulk
surface, the introduction of the substrate may bring the significant
effect upon the values of the barrier height. 
However, it is pausible to expect that the hierarchy in the barrier 
height of the saddle points is not much altered\cite{ssi3}.   
If so, by improving the time resolution of the experiment, the complicated 
simultaneous rearrangement of surface atoms illustrated above 
should be experimentally observed.
%%%%%%%%%%%%%%%%%%%%%%%%%%%%%%%%%%%%%%%%%%%
%  Fig.4
%%%%%%%%%%%%%%%%%%%%%%%%%%%%%%%%%%%%%%%%%%%
\begin{figure}[htbp]
\begin{center}
\resizebox{9.6cm}{!}{\includegraphics{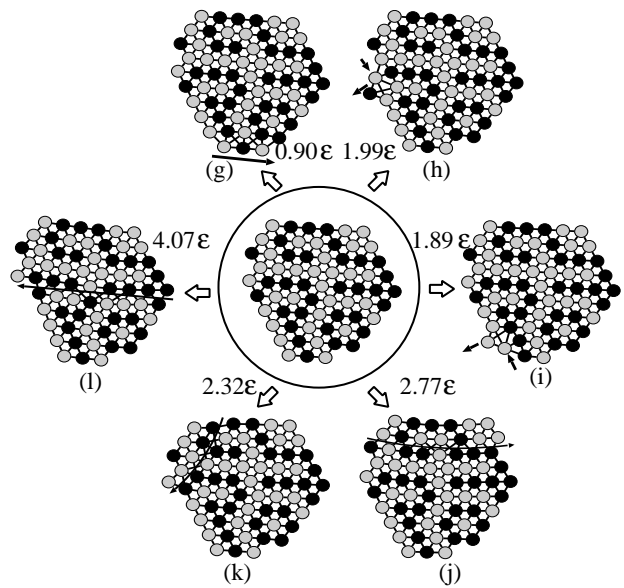}}
\caption{
Representative configurations of the cluster atoms corresponding to the 
1st order saddle points for the compact $M_{67}$ are exhibited by balls 
and sticks in (g)-(l). 
\label{fig:fig4}}
\end{center}
\end{figure}
%%%%%%%%%%%%%%%%%%%%%%%%%%%%%%%%%%%%%%%%%%%
%\subsection{Saddles with the high barrier height and their corresponding motion}
However, the isomerization process, where atoms located deep inside of 
the clusters are rearranged as well as surface atoms, yields the
apparently high energies barrier of saddle points as depicted 
in Fig.3(f) and Fig.4(l). These motions induced by a running crack
are infrequent event and are expected to be hard to occur, when the temperature
of the system is well below the melting point.     
%%%%%%%%%%%%%%%%%%%%%%%%%%%%%%%%%%%%%%%%%%%
%  Fig.5
%%%%%%%%%%%%%%%%%%%%%%%%%%%%%%%%%%%%%%%%%%%
\begin{figure}[htbp]
\begin{center}
\resizebox{9.5cm}{!}{\includegraphics{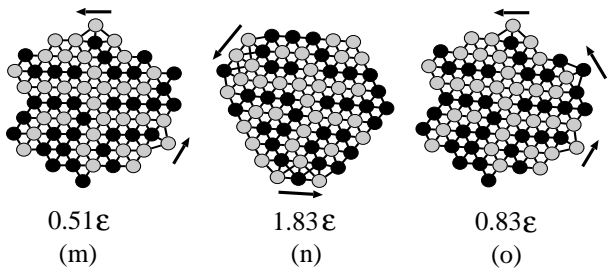}}
\caption{
Representative configurations of atoms corresponding to the higher-index saddle points for 
the non-compact and the compact $M_{67}$ are exhibited by balls and sticks. 
(m) and (o) reveal the 2nd and 3rd order saddles of the non-compact 
$M_{67}$, respectively. (n)is a manifestation of the 2nd order saddle
of the compact $M_{67}$.
\label{fig:fig5}}
\end{center}
\end{figure}
%%%%%%%%%%%%%%%%%%%%%%%%%%%%%%%%%%%%%%%%%%%%%%%%%%%%%%%%%%%%%%%%%%
%\section{Conclusion}
%{\it Conclusion}-
\par
In summary, plenty of saddles on a multidimensional PES of the 2D Morse clusters,
are numerically located in order to clarify the relation 
between the characteristic motion of surface atoms  
with the reaction path.
We evaluate the barrier height corresponding to 
various characteristic atomic motion for compact and non-compact $M_{67}$.
Our findings are: 
%%%%%%%%%%%%%%%%%%%%%%%%%%%%%%%%%%%%%%%%%%%%%%%%%%%%%%%%%%%%%%
(i)The 1st order saddles implying the lowest barrier energy is 
an indication of a {\it gliding motion} of a {\it floater}.
In particular, a {\it collective gliding motion} of a train 
composed of a few surface atoms also appears as low-lying saddles on PES. 
As the combination of these two types of the low-lying 1st order
saddles, there appear many low-lying higher-order saddles, which 
represent the {\it simultaneous gliding motion} of {\it
floaters} especially in the non-compact cluster.  
Note that the usual highder order saddles do not neccessarily 
correspond to the {\it simultaneous gliding} motions which are 
spatially separated on the cluster surface. 
%If an initial shape of a cluster is non-compact and possessed by floaters, 
%the barrier height of the higher-index saddles are especially 
%lowered. Most of them have their origin in  
%the {\it simultaneous gliding motion} of floaters.
(ii)Another characteristic atomic motion which implies sufficiently low  
barrier energy is associated with the collective behavior of surface
atoms, where {\it floaters} are created or annihilated. 
%accompanied with not only {\it floaters} but also surface atoms.
The saddle point for such a motion lies relatively high, but  
is still reachable at substantially low temperature 
below the melting point.
(iii)On the other hand, atomic displacements, where atoms located deep 
inside of the clusters are rearranged as well as surface atoms, yield the
high-lying saddle points which are scarcely reachable below the melting temperature. 
From these observations, we emphasize that the motion illustrated in (i)
and (ii) are expected to be observed in rearrangements of a surface
cluster, provided that the time resolution of the the experimental 
technique is improved. 
%\par
The above-mentioned ingredients on the PES of a non-compact and a compact
clusters lead us to the following overview about how the isomerization 
of a generic cluster proceeds in its dynamics substantially below the 
melting point: 
When the cluster shape becomes non-compact, {\it floaters}  and the train of
surface atoms actively migrate to form a resultant stable compact
shape. 
The quiescent state of the compact cluster will last for relatively
longer period until the atomic rearrangement  
eventually generates new {\it floaters} by climbing over the saddles which 
are exemplified in Fig.4(h) and (i). 
If there appear {\it floaters} and train of surface atoms in this way, an 
active surface motion again continues to alter the cluster shape to be compact.
A cluster keeps changing its shape from compact to non-compact
intermittently by activating and suppressing motion of the surface atoms.\par
%%%%%%%%%%%%%%%%%%%%%%%%%%%%%%%%%%%%%%%
Authors acknowledge Dr.T.Kobayashi and Prof.H.Yasuda for their helpful
comments.
One of authors(Y.S) thanks Prof.K.Hirao for his encouragement and 
the financial support from NEDO fellowship.
The present work was partly supported by the 
fellowship of Kwansei Gakuin University.
%%%%%%%%%%%%%%%%%%%%%%%%%%%%%%%%%%%%%%%

%%%%%%%%%%%%%%%%%%%%%%%%%%%%%%%%%%%%%%%
%\end{multicols}

\begin{thebibliography}{99}
\bibitem{note2}
The melting point of $M_{67}$ is about 670K($0.168\epsilon$), which is numerically
determined by locating an abrupt jump in the caloric curve.
\end{thebibliography}
\end{document}